\documentclass[superscriptaddress,prl,twocolumn,nofootinbib]{revtex4}

\usepackage{amsmath,amssymb}
\usepackage{dcolumn,array}

\usepackage{epsfig,psfrag}

\usepackage{mathrsfs}
\newcommand{\mrs}{\mathscr}

\begin{document}

  \title{Novel $NN$ interaction and the spectroscopy of light nuclei}
\author{A. M. Shirokov}
\affiliation{ Skobeltsyn Institute of Nuclear Physics, Moscow State University,
Moscow, 119992, Russia}
\author{J. P. Vary}
\affiliation{Department of Physics and Astronomy,
Iowa State University, Ames, Ia 50011-3160, USA}
\author{A. I. Mazur}
\affiliation{Physics Department, Khabarovsk State Technical University,
Tikhookeanskaya 136, Khabarovsk 680035, Russia}
\author{ S. A. Zaytsev}
\affiliation{Physics Department, Khabarovsk State Technical University,
Tikhookeanskaya 136, Khabarovsk 680035, Russia}
\author{T. A. Weber}
\affiliation{Department of Physics and Astronomy,
Iowa State University, Ames, Ia 50011-3160, USA}
  \begin{abstract}
  Nucleon-nucleon ($NN$) phase shifts and the spectroscopy of $A \le 6$
  nuclei are successfully described by an inverse scattering potential
  that is separable with oscillator form factors.
\end{abstract}
\maketitle


  Nucleon-nucleon ($NN$) potentials that describe available two-body
  data have a long and multi-faceted history.   High precision fits
  have improved with time even as more precise experimental data have
  become available. Three-nucleon ($NNN$) potentials have a shorter
  history but are intensively investigated at the present time.
  Disparate foundations for these potentials, both $NN$ \& $NNN$, have
  emerged.  On the one hand, one sees the predominant meson-exchange
  potentials sometimes supplemented with phenomenological terms to
  achieve high accuracy in fitting $NN$ data (Bonn \cite{Bonn},
  Nijmegen \cite{Stocks}, Argonne \cite{Argonne}, 
Idaho \cite{Idaho}, IS \cite{Plessas})
and $NNN$ data 
(Urbana \cite{Urbana,GFMC}, {Illinois} \cite{Illinois},
Tucson--Melbourne \cite{TM,TM-pr}). On
  the other hand, one sees the emergence of potentials with ties to
  QCD which are either meson-free \cite{vanKolck}, or intertwined
  with meson-exchange theory \cite{Bochum,Idaho}.

  All these potentials are being used, with
  unprecedented success, to explain a vast amount of data on light
  nuclei in Quantum Monte Carlo approaches \cite{GFMC} and  {\it ab initio}
  no-core shell model (NCSM) \cite{Vary,Vary3}. 
The overwhelming success of these
  efforts have led some to characterize these approaches as leading to
  a `Standard Model' of non-relativistic nuclear physics.

  Chief among the outstanding
  challenges is the computational intensity of using these $NN + NNN$
  potentials within the presently available many-body methods. For
  this reason, most {\it ab initio} investigations have been limited
  to $A \le 12$. The situation would be dramatically simpler if either
  the $NN$ potential alone would be sufficient or the potentials would
  couple less strongly between the low momentum and the high momentum
  degrees of freedom. If both simplifications are obtained, the future
  for applications is far more promising.

  In the present work, we derive and apply a new class of potentials
  which have no apparent connection with the two well-established
  lines of endeavor.  We develop
$J$-matrix inverse
  scattering potentials (JISP)
that describe $NN$ data to high accuracy
  and, with the off-shell freedom that remains, we obtain excellent
  fits to the bound and resonance states of light nuclei up to $A = 6$.
  Our $NN$ off-shell freedom is sufficient to describe
  these limited data  without the need for
  $NNN$ potentials.  As an important side benefit,
  we find that these potentials lead to rapid convergence in
  the {\it ab initio}
  NCSM evaluations presented here.
  We hope that these potentials will
  open a fruitful path for evaluating heavier systems and
  spur the development of extensions
  to scattering problems.

  It is important to stress that our $NN$ potentials have the same
  symmetries as the conventional $NN$ potentials mentioned above
  (without charge symmetry breaking at present), but they are not
  constrained by meson exchange theory, by QCD or by locality.  This
  does not mean our $NN$ potentials are inconsistent with those
  constraints, however.

By means of the $J$-matrix inverse scattering approach \cite{ISTP}
we construct $NN$ potentials as matrices in an oscillator basis
with $\hbar\omega=40$ MeV using the
Nijmegen $np$ phase shifts \cite{NNonline}.
Following Ref. \cite{ISTP}, we obtain inverse scattering tridiagonal
potentials (ISTP) that are tridiagonal (quasi-tridiagonal)
in uncoupled (coupled) partial waves.  The dimension of the
potential matrix is specified by the maximum value of $N=2n+l$
and is referred to as an $N\hbar\omega$
potential. In order to improve
the description of the phase
shifts, we develop a $9\hbar\omega$-ISTP in odd
waves instead of the $7\hbar\omega$-ISTP of
Ref. \cite{ISTP}. We retain an $8\hbar\omega$-ISTP in the
even partial waves.


Next we perform various phase equivalent transformations (PETs) of the
obtained ISTP. In the coupled $sd$ waves, we perform the same PET as
in Ref. \cite{ISTP} but with different rotation angle
$\vartheta=11.3^\circ$ to improve the description of the deuteron
quadrupole moment $Q$. We then find improvement in $^3$H and $^4$He
binding energies. We also perform similar PETs mixing lowest oscillator basis
states in the $^3p_2$, $^3p_1$, $^3d_2$ and $^1p_1$ waves with
the rotation angles of $\vartheta=+8^\circ$, $-6^\circ$, $+25^\circ$ and
$-16^\circ$  respectively to improve the description of the $^6$Li
spectrum. The obtained interaction fitted to the spectrum of $A=6$ nuclei, is
refered to as JISP6.
The non-zero matrix elements of the  JISP6 interaction
are presented in Tables~\ref{pot1s0}--\ref{pot3pf1} (in
$\hbar\omega=40$~MeV units).

\extrarowheight=3pt
\begin{table}
\caption{Non-zero matrix elements in $\hbar\omega=40$~MeV units  of the
JISP6 matrix in the $^1s_0$ partial wave.}
\begin{ruledtabular}
\begin{tabular}{>{$}c<{$}>{$}r<{$}>{$}r<{$}} 
n & \multicolumn{1}{c}{$ V_{nn}^l$}  & 
\multicolumn{1}{c}{$V_{n,\,n+1}^l=V_{n+1,\,n}^l$} 
\\ \hline
  0& -0.370829835410    & 0.132663053234 \\
  1&  -0.148826473896  &  0.006448104375 \\
  2&  0.152835073179   &  -0.120193538281   \\
  3&  0.187138532148   &  -0.029504403821 \\
  4& -0.005584124194  & \\ 
\end{tabular}
\label{pot1s0}

\caption{Non-zero matrix elements in $\hbar\omega=40$~MeV units  of the
JISP6 matrix in the $^1p_1$ partial wave.}
\begin{tabular}{>{$}c<{$}>{$}r<{$}>{$}r<{$}>{$}r<{$}} 
  n &  \multicolumn{1}{c}{$V_{nn}^l$}
         & \multicolumn{1}{c}{$ V_{n,\,n+1}^l=V_{n+1,\,n}^l$}
         & \multicolumn{1}{c}{$ V_{n,\,n+2}^l=V_{n+2,\,n}^l$}   \\ \hline
  0& 0.631081576474     &  -0.251382936855  &  0.413319237868   \\
  1& -0.293390247307    &  -0.118539824524     \\
  2&  0.454133632867     &  -0.230186013455  \\
  3&  0.348035837559  &     -0.090043227032 \\
  4&  0.049221178231 
\end{tabular}\label{pot1p1}

  \caption{Non-zero matrix elements in $\hbar\omega=40$~MeV units  of the
JISP6 matrix in the $^1d_2$ partial wave.}
\begin{tabular}{>{$}c<{$}>{$}r<{$}>{$}r<{$}} 
  n &  \multicolumn{1}{c}{$V_{nn}^l$}
        & \multicolumn{1}{c}{$ V_{n,\,n+1}^l=V_{n+1,\,n}^l$}   \\ \hline
  0  &-0.040699390045    &   0.037531685348    \\
  1  &  -0.111761745807   &  0.069791608541  \\
  2  & -0.134996618188    &  0.045265020638   \\
  3  & -0.031270631267 &    \\ 
\end{tabular}\label{pot1d2}

\caption{Non-zero matrix elements in $\hbar\omega=40$~MeV units  of the
JISP6 matrix in the $^1f_3$ partial wave.}
\begin{tabular}{>{$}c<{$}>{$}r<{$}>{$}r<{$}} 
  n &  \multicolumn{1}{c}{$V_{nn}^l$}
         & \multicolumn{1}{c}{$ V_{n,\,n+1}^l=V_{n+1,\,n}^l$}   \\ \hline
  0&  0.019468932284    &-0.018631285428   \\
  1&  0.083569595544   & -0.055488297878   \\
  2&  0.106821875609   & -0.032273326919   \\ 
  3&  0.021063860199
\end{tabular}\label{pot1f3}
\end{ruledtabular}
\end{table}

\begin{table}
\caption{Non-zero matrix elements in $\hbar\omega=40$~MeV units  of the
JISP6 matrix in the $^3p_0$ partial wave.}
\begin{ruledtabular}
\begin{tabular}{>{$}c<{$}>{$}r<{$}>{$}r<{$}} 
  n &  \multicolumn{1}{c}{$V_{nn}^l$}
        & \multicolumn{1}{c}{$V_{n,\,n+1}^l=V_{n+1,\,n}^l$}   \\ \hline
  0& -0.143164548564    & 0.020755069079    \\
  1&  0.082988173615    & -0.120094506156    \\
  2&  0.310447079465   &  -0.116102071897    \\
  3&  0.065044984944     & 0.013609203897 \\ 
  4&  -0.026555044004
\end{tabular}\label{pot3p0}

  \caption{Non-zero matrix elements in $\hbar\omega=40$~MeV units  of the
JISP6 matrix in the $^3p_1$ partial wave.}
\begin{tabular}{>{$}c<{$}>{$}r<{$}>{$}r<{$}>{$}r<{$}}
  n & \multicolumn{1}{c}{$V_{nn}^l$}
         &\multicolumn{1}{c}{$ V_{n,\,n+1}^l=V_{n+1,\,n}^l$}
         &\multicolumn{1}{c}{$ V_{n,\,n+2}^l=V_{n+2,\,n}^l$}   \\ \hline
  0& 0.249679784904   &  -0.164761352606 & 0.157602869176   \\
  1&  0.044327922667   & -0.176615480839   \\
  2&  0.514099248363   &  -0.275733929941  \\
  3&  0.423324941377   &  -0.108223480414 \\  
  4&   0.055397268090
\end{tabular}\label{pot3p1}

\caption{Non-zero matrix elements in $\hbar\omega=40$~MeV units  of the
JISP6 matrix in the $^3d_2$ partial wave. }
\begin{tabular}{>{$}c<{$}>{$}r<{$}>{$}r<{$}>{$}r<{$}} 
  n & \multicolumn{1}{c}{$V_{nn}^l$}
    & \multicolumn{1}{c}{$ V_{n,\,n+1}^l=V_{n+1,\,n}^l$}
    & \multicolumn{1}{c}{$ V_{n,\,n+2}^l=V_{n+2,\,n}^l$}   \\ \hline
  0  & -0.662113235725   &  0.759732268972  & -0.571851583929  \\
  1  & 0.175448232529   &  0.273660320760   \\
  2  & -0.263880156721  & 0.086029122705    \\
  3  & -0.063231092747  &    \\
\end{tabular}
\label{pot3d2}

  \caption{Non-zero matrix elements in $\hbar\omega=40$~MeV units  of the
JISP6 matrix in the $^3f_3$ partial wave.}
\begin{tabular}{>{$}c<{$}>{$}r<{$}>{$}r<{$}} 
  n &  \multicolumn{1}{c}{$V_{nn}^l $}
       &  \multicolumn{1}{c}{$V_{n,\,n+1}^l=V_{n+1,\,n}^l $}  \\ \hline
  0&  0.026326206898    &  -0.014285757490 \\
  1&  0.035674429367   &  -0.016797566427   \\
  2&  0.028543592124  & -0.008290586001   \\ 
  3& 0.006036946613
\end{tabular}\label{pot3f3}
\end{ruledtabular}
\end{table}

\begin{table}
\caption{Non-zero matrix elements  in $\hbar\omega=40$~MeV units of
the JISP6 matrix in the $sd$ coupled waves.} \label{potsd1}
\begin{ruledtabular}
\begin{tabular}{>{$}c<{$}>{$}r<{$}>{$}r<{$}>{$}r<{$}} 
  \multicolumn{3}{c}{$V^{ss}_{nn'}$ matrix elements} \\
   n &  \multicolumn{1}{c}{$V_{nn}^{ss}$}
       &\multicolumn{1}{c}{$ 
V_{n,n+1}^{ss}=V^{ss}_{n+1,n}$}  \\ \cline{1-3} 
  0& -0.508274040822 &  0.214156446587   \\
  1&  -0.276168029473     & 0.080907735691  \\
  2&  -0.009473803659   & -0.051881443108  \\
  3&  0.152873734289   & -0.055193589842  \\
  4&   0.037547929880 &  \\[1.5ex] 
  \multicolumn{3}{c}{$V^{dd}_{nn'}$ matrix elements} \\  
  n &\multicolumn{1}{c}{$ V_{nn}^{dd}$}
         & 
\multicolumn{1}{c}{$V_{n,n+1}^{dd}=V^{dd}_{n+1,n}$} 
\\ \cline{1-3} 
0&   0.050878349132    &  -0.094173649477  \\
1 &  0.322126471805    & -0.178808793641   \\
2&   0.308516673061    & -0.093012604766   \\
3&  0.061200037193  &\\[1.5ex] 
   \multicolumn{4}{c}{$V^{sd}_{nn'}=V^{ds}_{n'n}$ matrix elements} \\  
   n &\multicolumn{1}{c}{$ V_{n,n-1}^{sd}=V^{ds}_{n-1,n}$}
        &\multicolumn{1}{c}{$ V_{nn}^{sd}=V_{nn}^{ds}$}
     & \multicolumn{1}{c}{$ V_{n,n+1}^{sd}=V^{ds}_{n+1,n}$} \\ \hline
  0&     & -0.411771355351   &  0.205731982741   \\
  1&   -0.048516342749    & -0.060476585693  &  \\
  2&    0.068044496963   &  -0.080187106458  &  \\
  3&    0.049400578816     &  -0.020205646231    & \\
  4&  -0.001503998139  &  &   
\end{tabular}
\end{ruledtabular}
\end{table}

\begin{table} 
\caption{Non-zero matrix elements in $\hbar\omega=40$~MeV units of the
JISP6 matrix in the $pf$ coupled partial waves.}
\label{pot3pf1}
\begin{ruledtabular}
\begin{tabular}{>{$}c<{$}>{$}r<{$}>{$}r<{$}>{$}r<{$}}
  \multicolumn{4}{c}{$V^{pp}_{nn'\vphantom{,}}$ matrix elements} \\
   n &\multicolumn{1}{c}{$ V_{n n}^{pp}$}
    & \multicolumn{1}{c}{$V_{n, n+1}^{pp}=V_{n+1,n}^{pp}$}
    & \multicolumn{1}{c}{$V_{n, n+2}^{pp}=V_{n+2,n}^{pp}$}\\ \hline
  0& -0.257052769018   & 0.215269922150 &  -0.171332097424 \\
  1&   0.035950531483  &  0.103784437050  \\
  2&  -0.209221226255    & 0.103221679634   \\
  3&  -0.151546344031  &  0.037329967115 \\
  4&  -0.015878201100  \\[1.5ex] 
  \multicolumn{3}{c}{$V^{ff}_{nn'\vphantom{,}}$ matrix elements} \\
n & \multicolumn{1}{c}{$V_{n n}^{ff}$}
        & \multicolumn{1}{c}{$V_{n, n+1}^{ff}= V_{n+1,n}^{ff}$}
        & 
          \\ \cline{1-3} 
0&  -0.019836117401  &  0.008292672214  \\
1&   -0.010058323807   &  0.000628665253   \\
2&  0.001646202464  & -0.000973797731 \\
3&  0.000388504318 \\[1.5ex] 
   \multicolumn{4}{c}{ $V^{pf}_{nn'\vphantom{,}}$ matrix elements } \\
   n &  \multicolumn{1}{c}{$V_{n, n-1}^{pf}= V_{n-1,n}^{fp}$}
      & \multicolumn{1}{c}{$ V_{n n}^{pf}=V^{fp}_{nn}$}
      & \multicolumn{1}{c}{$ V_{n, n+1}^{pf}=V^{fp}_{n+1, n}$} \\ \hline
  0&     & 0.018163699530 &  0.003299666393   \\
  1&  -0.026186898553  &   0.023478346345   \\
  2&  -0.024757588981   &  0.023707438623    \\
  3&    -0.014708906826   &   0.006271279847 \\
  4&   0.000024653107  
\end{tabular}
\end{ruledtabular}
\end{table}

The deuteron properties
provided by JISP6 are compared with those 
of some other
realistic potentials in Table \ref{d-prop}.

\begin{table*}
\caption{JISP6 deuteron property predictions  in
comparison with the ones obtained with various realistic potentials.}
\label{d-prop}
\begin{ruledtabular}
\begin{tabular}{ccccccc} 
Potential & $E_{d}$, MeV &\parbox{2.1cm}{$d$ state probability, \%}
& \parbox{1.6cm}{rms radius, fm} & $Q$, fm$^2$
                    & \parbox{2.4cm}{As. norm. const.
                    ${\mrs A}_s$, fm$^{-1/2}$}
&$\displaystyle\eta=\frac{{\mrs A}\vphantom{'}_d}
           {{\mrs A}_{s\vphantom{_a}}}$\\ \hline
JISP6 & $-2.224575$ & 4.1360 & 1.9647 &0.2915 &0.8629 &0.0252\\
  Nijmegen-II & $-2.224575$ & 5.635 & 1.968 & 0.2707 &0.8845 & 0.0252\\
AV18 & $-2.224575$ &5.76 & 1.967 & 0.270 & 0.8850 & 0.0250\\
CD--Bonn & $-2.224575$ & 4.85 & 1.966 & 0.270 & 0.8846 &0.0256\\
Nature 
& $-2.224575(9)$ & --- 
&1.971(6)
&0.2859(3)  &0.8846(9) &0.0256(4)         
\end{tabular}
\end{ruledtabular}
\end{table*}

We perform calculations of light nuclei 
in the NCSM with
JISP6 plus the Coulomb interaction between protons. To improve the
convergence, we perform the Lee--Suzuki transformation to obtain a
two-body effective interaction as is discussed in
Ref. \cite{Vary3}.
We obtain the effective interaction in a new basis
($\hbar\omega=15$ MeV) within an
$N_{max}\hbar\omega$ model space where $N_{max}$ signifies the
many-body oscillator basis cutoff.
The results of our NCSM calculations for binding energies
of $^3$H,$^3$He (in the $14\hbar\omega$ model space), $^4$He (in the
$12\hbar\omega$ model space), $^6$He (in the $8\hbar\omega$ model space)
and $^6$Li (in the $10\hbar\omega$ model space) nuclei are compared in
Table \ref{bind}
with the calculations in various approaches [Faddeev, Green's-function
Monte Carlo (GFMC), NCSM] with realistic $NN$ [CD-Bonn,
Nijmegen-I (NijmI), Nijmegen-II (NijmII),  and
Argonne (AV18 and AV8')]  and $NNN$ [Urbana (UrbIX) and Tucson--Melbourne (TM
and TM')] potentials. To give an estimate of the convergence of our
calculations, we present the difference between the given result and
the result obtained in the next smaller model space
in parenthesis after our JISP6 results. It is seen that the convergence of
our calculations is adequate.

\begingroup
\begin{table*}
\extrarowheight=3pt
\begin{ruledtabular}
\caption{The binding energies of $^3$H,$^3$He, $^4$He, $^6$He and
$^6$Li nuclei. }
\label{bind}
\begin{tabular}{cccccc}
Potential &$^3$H &$^3$He &$^4$He &$^6$He &$^6$Li\\ \hline
JISP6, NCSM &8.461(5) &7.751(3) &28.611(41)&29.072(69) &31.48(27)\\
{{CD-Bonn+TM}, Faddeev \cite{alpha}} &8.480 &7.734 &29.15\\
{AV18+TM, Faddeev \cite{alpha}} &8.476 &7.756 & 28.84 \\
{AV18+TM, Faddeev \cite{alpha}} &8.444 &7.728 &28.36\\
{NijmI+TM, Faddeev \cite{alpha}} &8.392 &7.720 &28.60\\
{NijmII+TM, Faddeev \cite{alpha}} &8.386 &7.720 &28.54\\
{AV18+UrbIX, Faddeev \cite{alpha}} &8.478 &7.760 &28.50\\
{AV18+UrbIX, GFMC \cite{GFMC}} &8.47(1) & &28.30(2)
                        &27.64(14) &31.25(11)\\
{AV8'+TM', NCSM \cite{NaO}} & & & &28.189 & 31.036\\
Nature &8.48 &7.72 & 28.30 &29.269 &31.995
\end{tabular}
\end{ruledtabular}
\end{table*}
\endgroup

\begin{figure}
\centerline{\epsfig{file=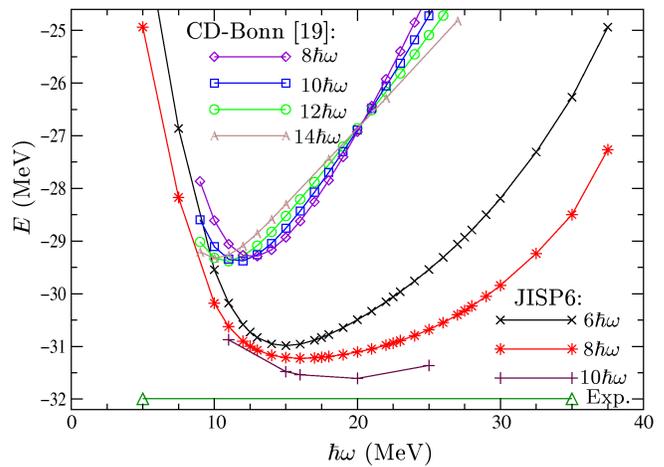,width=.48\textwidth}}
\caption{(Color online) $\hbar\omega$ dependence of the $^6$Li ground
state energy 
obtained with JISP6 interaction in comparison with the one obtained
in NCSM with CD-Bonn potential \protect\cite{NVOB}.}
\label{hwdep}
\end{figure}

The convergence patterns are also illustrated by Fig.~\ref{hwdep}
where we present the $\hbar\omega$ dependence of the $^6$Li ground
state energy in comparison with the results of Ref. \cite{NVOB}
obtained in NCSM with CD-Bonn interaction. The
$\hbar\omega$ dependence with the JISP6 interaction is weaker
over a wide interval of $\hbar\omega$ values. This is
a signal that convergence is improved relative to CD-Bonn.
The variational principle cannot be applied to
the NCSM calculations with effective interactions so the
convergence may be either from above or below.
However, we may surmise that the residual contributions of neglected
three-body effective interactions is  more significant in the
CD-Bonn case.


Returning to the results presented in
Tables~\ref{d-prop}--\ref{bind}, we see that the JISP6 interaction
provides a realistic description of
the ground states  of light nuclei competitive with the
quality of descriptions previously achieved with both $NN$
and $NNN$ forces.

This conclusion is supported by the
calculations of the  spectra of $A=6$ nuclei with $\hbar\omega=15$ MeV
presented in Table \ref{6LiSp}. We again present in parenthesis  the
difference between the given excitation energy and the result obtained
in the next smaller model space. It is seen that the $^6$Li
spectrum is well-reproduced in our calculations. The most important
difference with the experiment is the excitation energy of the
$(1^+_2,0)$ state. However $E_x(1^+_2,0)$ goes down rapidly when the
model space is increased and better results are anticipated
in a larger model space. The JISP6 results for
$^6$Li spectrum  are also seen to be competitive with results from
modern realistic $NN+NNN$ interaction models.  We note here that the
$^6$Li spectrum was found \cite{NaO} to be significantly sensitive to the
presence of the $NNN$ force and this motivated our adoption of $^6$Li
for these comparisons.

\begin{table}
\extrarowheight=3pt
\begin{ruledtabular}
\caption{Excitation energies
$E_x$ (in MeV) of $A=6$ nuclei.}
\label{6LiSp}
\begin{tabular}{ccccc}
$^6$Li &Nature &JISP6 & AV8'+TM' &\!AV18+UrbIX \\
\!Model space & & $10\hbar\omega$ &\!NCSM, $6\hbar\omega$ \cite{NaO}
                            &GFMC \cite{GFMC}\\ \hline
$E_x(1^+_1,0)$ &0.0 &0.0  &0.0     &0.0\\
$E_x(3^+,0)$ &2.186  &2.102(4) &2.471 &2.72(36) \\
$E_x(0^+,1)$ &3.563  &3.348(24) &3.886 &3.94(23) \\
$E_x(2^+,0)$ &4.312  &4.642(2) &5.010 &4.43(39)  \\
$E_x(2^+,1)$ &5.366  &5.820(4) &6.482   \\
$E_x(1^+_2,0)$ &5.65 
                       &6.86(36) &7.621 \\ \hline
$^6$He &Nature &JISP6 & AV8'+TM' &\!AV18+UrbIX \\
\!Model space & & $8\hbar\omega$ &\!NCSM, $6\hbar\omega$ \cite{NaO}
                            &GFMC \cite{GFMC}\\ \hline
$E_x(0^+,1)$ &0.0 &0.0  &0.0     &0.0\\
$E_x(2^+,1)$ &1.8 
                 &2.59(13)  &2.598     &1.80(18)
\end{tabular}
\end{ruledtabular}
\end{table}

  We return to the underlying rationale for our approach and ask why
  it is conceivable that an $NN$ interaction alone may be as successful
  as the $NN + NNN$ potentials mentioned at the outset.  That this is
  feasible may be appreciated from the theorem of Polyzou and
  Gl\"ockle~\cite{Poly}.  They have shown that changing the off-shell
  properties of
  two-body potentials is equivalent to adding  many-body
  interactions.  This theorem coupled with our limited results suggests
  that our inverse scattering $NN$ potential plus off-shell
  modifications is roughly equivalent, for the observables so far
  investigated, to the successful $NN + NNN$ potential models.

  Clearly, more work will be needed to carry this to nuclei with
$A  \ge 7$ and see if the trend continues.  Based on the results
  presented, the additional off-shell freedoms remaining may well
  serve to continue this line of fitting properties for some time.
  When it eventually breaks down, $NNN$ potentials may
  be needed.

This work was supported in part
by 
RFBR grant
No~02-02-17316,
by US~DOE grant No~DE-FG-02~87ER40371 and by  US~NSF grant
No~PHY-007-1027.

\end{document}